\documentclass[a4paper]{article}

\usepackage{icrc2013}
\usepackage[english]{babel}

\usepackage{amssymb,amsmath}

\newcommand{\apj}{The Astrophysical Journal}
\newcommand{\aap}{Astronomy \& Astrophysics}

\title{The 1st {\em Fermi} LAT SNR Catalog: Constraining the Cosmic Ray Contribution}

\shorttitle{{\em Fermi} SNR Cat: Constraining CRs}

\authors{
T.~J. Brandt$^{1}$,
F. Acero$^{1,2}$,
F. de Palma$^{3}$,
J. W. Hewitt$^{1,4}$,
M. Renaud$^{5}$,
for the {\em Fermi} LAT Collaboration.
}

\afiliations{
$^1$ NASA/Goddard Space Flight Center  \\
$^2$ NASA Postdoctoral Program Fellow \\
$^3$ INFN Sezione di Bari, 70126 Italia \\
$^4$ CRESST/University of Maryland, Baltimore County, Baltimore, MD 21250, USA \\
$^5$ Laboratoire Univers et Particules de Montpellier, Montpellier, France \\
}

\email{theresa.j.brandt@nasa.gov}

\abstract{
Despite tantalizing evidence that supernova remnants (SNRs) are the source of Galactic cosmic rays (CRs), including the recent detection of a spectral signature of hadronic $\gamma$-ray emission from two SNRs, their origin in aggregate remains elusive. We address the long-standing question of Galactic CR nuclei origins using our statistically significant GeV SNR sample to estimate the contribution of SNRs to directly observed CRs. Interactions between CRs and ambient gas near the SNRs emit photons via pion decay at GeV energies, providing an in situ tracer for CRs otherwise measured directly with balloon-borne and satellite experiments near the Earth. To date, the {\em Fermi} LAT SNR Catalog has detected more than 50 SNRs and potential associations in classes with a variety of properties, yet all remain possible accelerators. We investigate the GeV and multiwavelength (MW) emission from SNRs to constrain their maximal contribution to observed Galactic CRs. Our work demonstrates the need for improvements to previously sufficient simple models describing the GeV and MW emission from these objects.
}


\begin{document}
\maketitle

\section{Introduction}
Direct measurements of cosmic ray (CR) energetics and composition combined with our understanding of high energy accelerators in the Galaxy have long suggested that supernova remnants (SNRs) are likely the source of Galactic CRs. Yet proof has for a long while remained elusive. With the advent of $\gamma$-ray telescopes with degree-scale spatial resolution in addition to good spectral resolution, we have made significant strides towards more firmly associating SNRs with CR acceleration. These have included individual SNR spectra tending towards being dominated by hadronic emission, where CRs interacting with the local medium emit $\gamma$-rays via $\pi^0$ decay (e.g. \cite{DaveThompsonReview}), as well as recent detection of proton acceleration through the low-energy $\pi^0$ cutoff \cite{lowEnergySNRs}. 

Yet while such individual results are necessary, they are not sufficient, for the direct data we would ultimately like to compare 
to is comprised of CRs from sources throughout the Galaxy. Thus, it is also necessary to show that the aggregate contribution from all sources can produce the observed CRs. In order to do so, we have leveraged several years' worth of \textit{Fermi} Large Area Telescope (LAT) survey data to study systematically all known Galactic SNRs, the majority of which are detected in the radio and compiled in \cite{green_cat}, as well as a few identifications from other wavelengths. Details of the analysis procedure are laid out in \cite{JacksProceeding}, along with a discussion of the implications of a radio-GeV flux correlation. As interstellar $\gamma$-ray emission is quite prevalent along the galactic plane, where the majority of SNRs lie, we also explore systematics related to the choice of interstellar emission model in \cite{IEM_ICRCProceeding, IEMFSympProc} to ensure that we will have the most robust results possible. These results point to a separation of SNRs into classes, notably those which are young and those which are interacting, often with molecular clouds~\cite{JacksProceeding}. 

\section{Particle Populations}

In \cite{JacksProceeding}, we showed that the synchrotron radio emission from high energy leptons tends to be correlated for interacting SNRs, suggesting a physical link, whereas the young SNRs showed more scatter. 

\subsection{Emission Mechanisms}

\begin{figure}[h]
\centering
\includegraphics[width=0.4\textwidth]{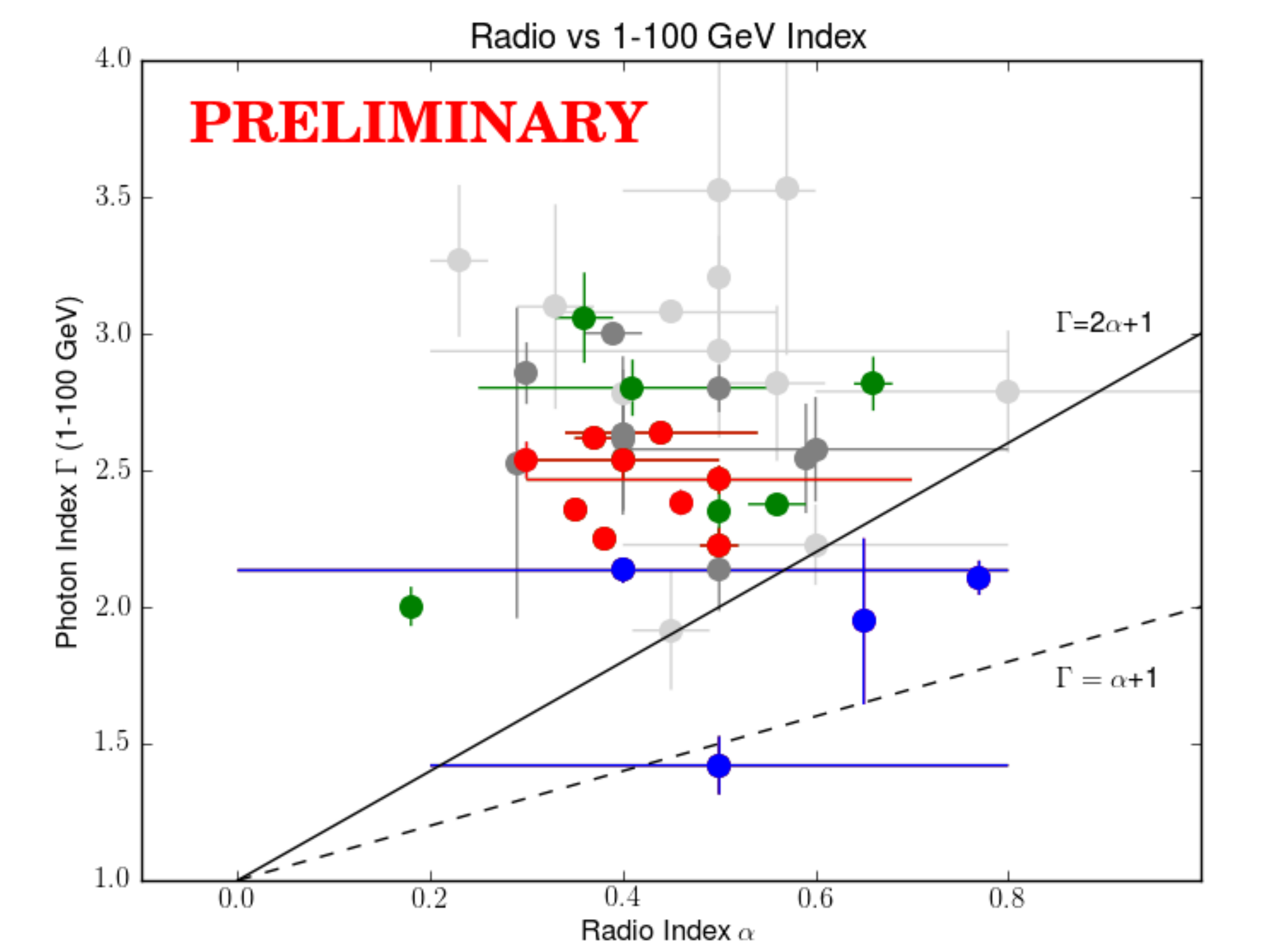}
\caption{GeV-Radio index and expected slope correlation for:  $\pi^0$ decay or e$^{\pm}$ bremsstrahlung (solid line) and inverse Compton (dashed line). As described in more detail in \cite{JacksProceeding}, the colors correspond to different types of SNRs and SNR candidates, namely, young SNRs are blue; those identified, interacting SNRs are red; the green points correspond to newly identified SNRs; and the grey points to point-like candidate SNRs (dark grey) and point-like candidate pulsars (light grey). This scheme is maintained throughout the paper.}
\label{fig:GeVradioIndex}
\end{figure}

If radio and GeV emission arise from the same particle population(s), e.g. leptons and hadrons accelerated at the SNR shock front, under simple assumptions, the GeV and radio indices should be correlated. For inverse Compton (IC) emitting leptons, the GeV and radio photon indices ($\Gamma$ and $\alpha$ respectively) can be related as $\Gamma = \alpha + 1$ whereas for $\pi^0$ decay and e$^\pm$ bremsstrahlung, $\Gamma = 2\alpha + 1$. Figure \ref{fig:GeVradioIndex} shows that, contrary to our radio/GeV flux observations, young SNRs seem consistent with expectations from these simple models. Several of the known, young SNRs are more consistent with the IC relation (dashed line), suggesting that they may be lepton-dominated and emitting via IC in the GeV regime. SNRs emitting via a combination of mechanisms in this scenario have indices falling between the two index relations, that is, the region spanned by the $\pi^0$/bremsstrahlung (solid) and IC (dashed) lines. The young SNR RX J1713-3946 is one example which bears out this case \cite{rxjFermi}. 
Other SNRs, including those observed to be interacting with molecular clouds, are softer than expected, and in most cases are not even consistent with combinations of emission mechanisms.

The apparent lack of correlation between the indices and emission mechanisms for the majority of observed SNRs suggests that the data are now able to challenge model assumptions for these SNRs. These assumptions include that: 
\begin{itemize}
\item the underlying leptonic and hadronic populations may have different power law indices; 
\item the emitting particle population(s) may not follow a power law but may instead have break(s); 
\item or there may be different zones with different properties dominating the emission at different wavelengths.
\end{itemize}

\subsection{Spectral Break?}

\begin{figure}[h]
  \centering
   \includegraphics[width=0.4\textwidth]{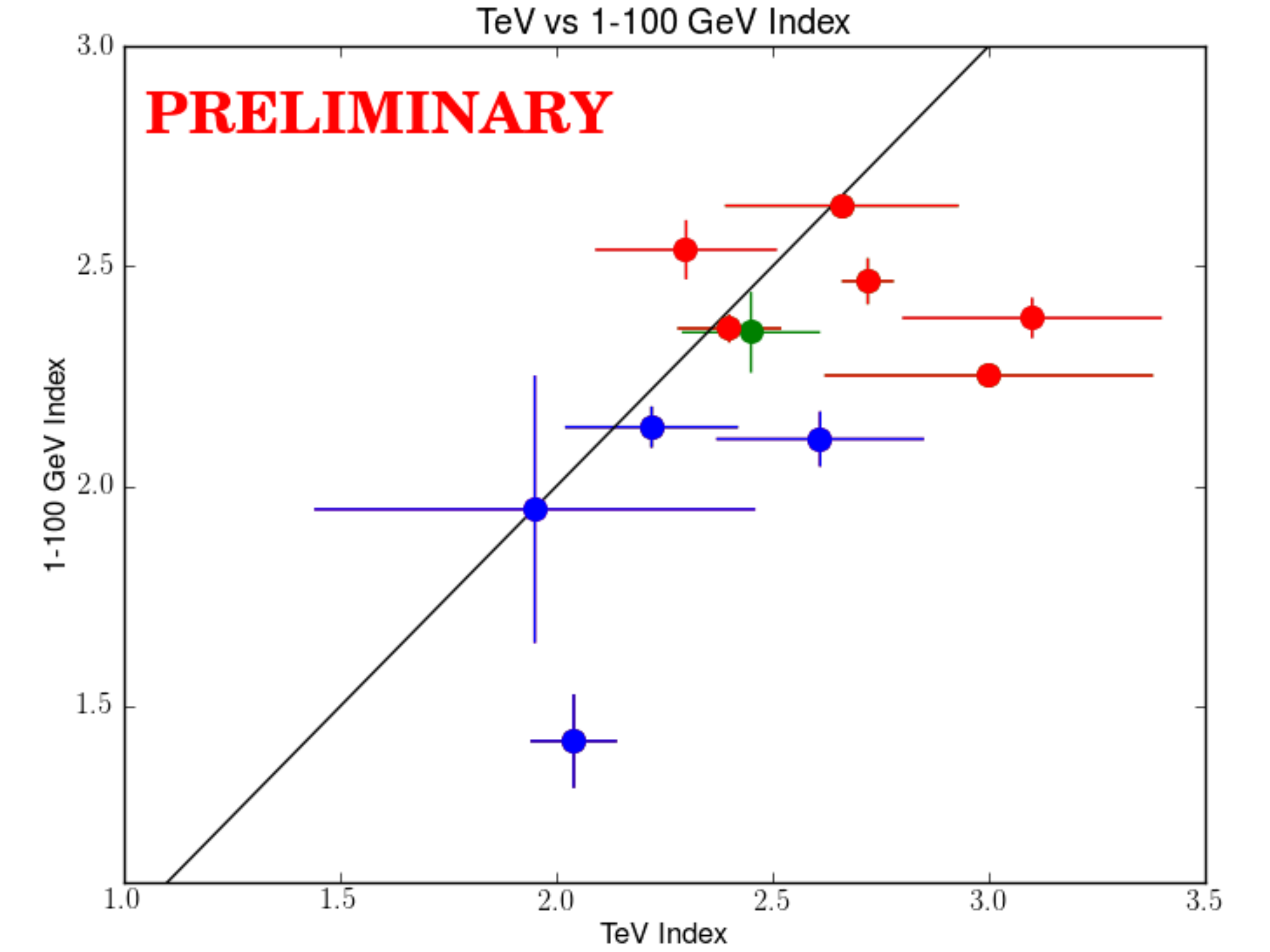}
  \caption{GeV-TeV index. The line shows equal indices. SNRs lying below the line suggest that their spectra have breaks, potentially reflecting a break in the underlying particle population(s') index or indices.}
  \label{fig:GeVTeVIndex}
 \end{figure}

With SNRs studied in TeV, we have the opportunity to explore the second of the model assumptions: that the emitting particle population(s) may have breaks.
Such a break in the underlying particle population(s) can also cause a break in the observed spectrum. As TeV emission may arise from the same mechanisms as the {\em Fermi}-observed GeV emission, we might expect to see such a break reflected in the spectrum combining {\em Fermi} data with observations from Imaging Air Cherenkov Telescopes (IACTs) such as H.E.S.S.,  VERITAS, and MAGIC. In Figure \ref{fig:GeVTeVIndex} we plot the GeV index versus TeV index for all SNRs observed with both  {\em Fermi} and an IACT. 
Several SNRs' TeV indices are lower than their GeV index, and few lie above the line of equal index. This suggests a break either at or between GeV and TeV energies. The former has been observed for e.g. IC443 \cite{IC443FermiTeV} and the latter in, e.g. the young SNR RX J1713-3946 \cite{rxjFermi}. Such a break would be a tantalizing clue, likely reflecting a break in the underlying particle spectrum, as it does for IC443 and RX J1713-3946, of many SNRs.
We note however that, as the TeV sources are not uniformly surveyed, inferring population statements from this observation requires a careful understanding of the non-TeV observed SNR subsample.

We note that the GeV-TeV index plot (Fig. \ref{fig:GeVTeVIndex}) also shows a distinct separation between young and interacting, often older SNRs, suggesting an evolution in index with age, from harder when younger to softer when older. We explore this further by explicitly investigating the evolution of the GeV index with age in the next section.

\section{Evolution or Environment?}

\begin{figure}[h!]
  \centering
  \includegraphics[width=0.4\textwidth]{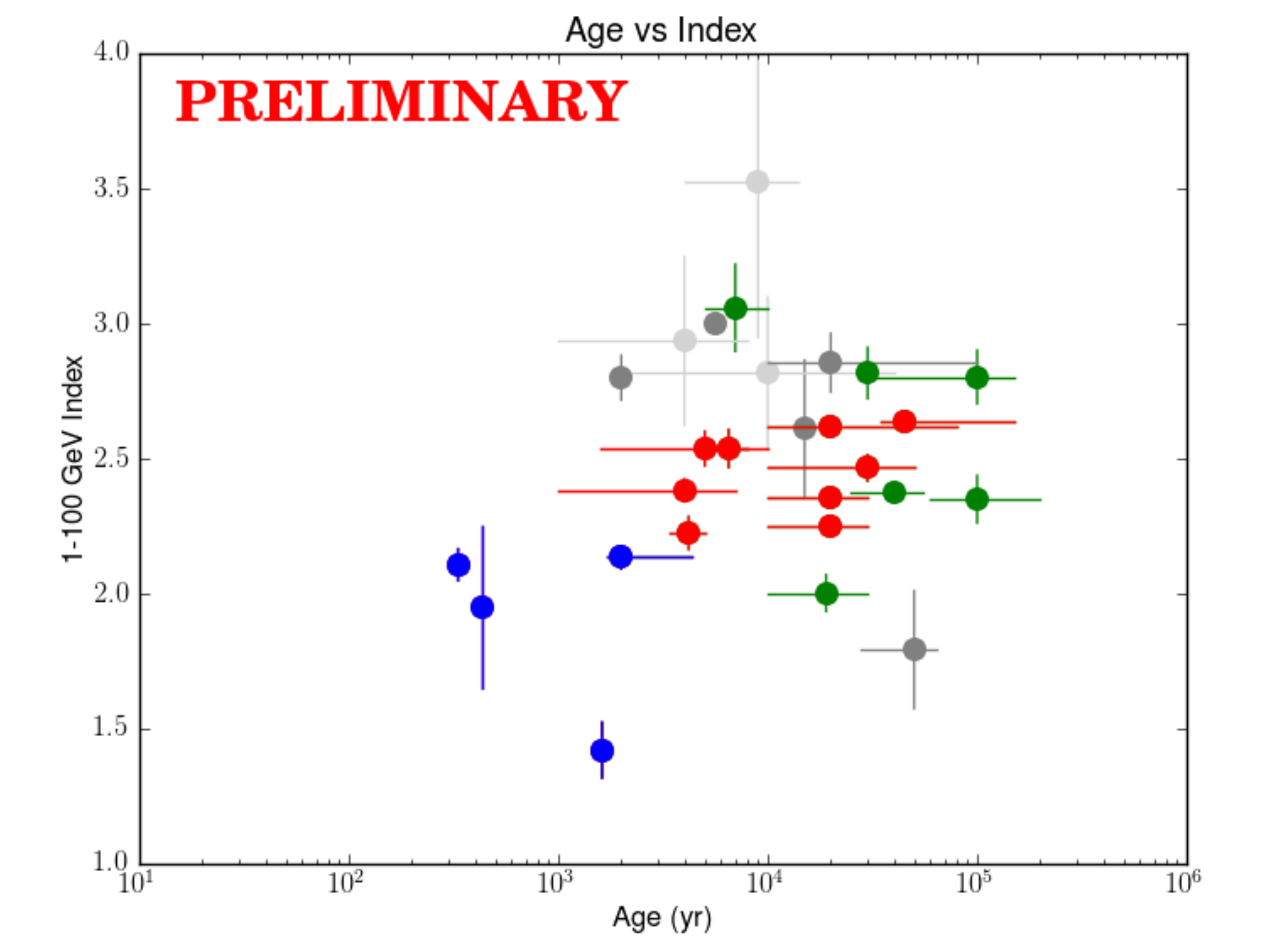}
  \caption{Age versus GeV index. The young (blue) SNRs are separated in GeV index from the identified interacting SNRs (red).
 }
  \label{fig:AgeGeVIndex}
 \end{figure}

As we saw a division between young SNRs having harder indices and interacting SNRs' tending to softer ones, 
we explicitly examine the evolution of the GeV index with age of the SNR. In Figure \ref{fig:AgeGeVIndex} we see a clear separation of known young SNRs having lower, harder GeV indices than interacting SNRs, where ages were drawn from the literature. 
This separation could be due to decreasing shock speed, decreasing the maximum acceleration energy as SNRs age. Particles will be more readily able to escape a slower shock, thereby reducing the $\gamma$-ray flux. 
Further, a less energetic shock will no longer be able to accelerate particles to the highest energy, thereby reducing emission at the highest energies. 

In \cite{JacksProceeding}, we observed that SNRs known to be interacting, in particular with large molecular clouds, appear to be more luminous in GeV $\gamma$-rays than young SNRs. 
The apparent difference in indices and luminosities for the young and interacting SNRs may also be caused by differing environments: 
older SNRs may interact with denser surroundings not yet reached by younger SNRs.
Using MW information such as ambient density estimated from thermal X-rays in the catalog context will help disentangle the effects of evolution and environment.

\section{Constraining CR Acceleration}
Recent work examining GeV $\gamma$-ray data at the $\pi^0$ rest mass \cite{lowEnergySNRs} has added another piece to the accumulating evidence (e.g. \cite{DaveThompsonReview, CTB37A_COSPAR, CastroSlane}) that SNRs accelerate hadrons: at least two SNRs, IC443 and W44, show evidence of the $\pi^0$ low energy break ($E<100$\,MeV), demonstrating that they accelerate protons. 
To this necessary evidence that SNRs accelerate hadrons, we must also add an understanding of the Galactic SNR population's ability to accelerate the appropriate composition of hadrons to the energies observed by direct-detection experiments. 

For a SNR at a given distance $d$ interacting with a density $n$ and accelerating cosmic rays to a maximum energy $E_{CR, max} \gtrsim 200$\,GeV with index $\Gamma_{CR} \approx 2.5$, we can relate the $\gamma$-ray flux above $1$\,GeV to the SNR's energy, $E_{SN}$, and CR acceleration efficiency, $\epsilon_{CR}\equiv\frac{E_{CR}}{E_{SN}} $, as: 
\begin{equation}
\begin{split}\label{eq:accelEfficiency}
F(>1\,\textrm{GeV})  \approx  10^{-8}  & \times \frac{\epsilon_{CR}}{0.1} \times \frac{E_{SN}}{10^{51}\,\textrm{erg}} \\
& \times \frac{n}{1\,\textrm{cm$^{-3}$}}  \times \left(\frac{d}{1\,\textrm{kpc}}\right)^{-2} \,\textrm{cm$^{-2}$ s$^{-1}$} 
\end{split}
\end{equation}
which is consistent with e.g. \cite{DruryAharonianVoelk1994}. It is useful to note that this derivation includes the approximation that the majority of the transfer of SN explosion energy to hadrons occurs during the Sedov(-like) phase, and that the efficiency remains roughly constant during this period (see \cite{DruryAharonianVoelk1994} for further discussion).

\begin{figure}[h]
  \centering
  \includegraphics[width=0.4\textwidth]{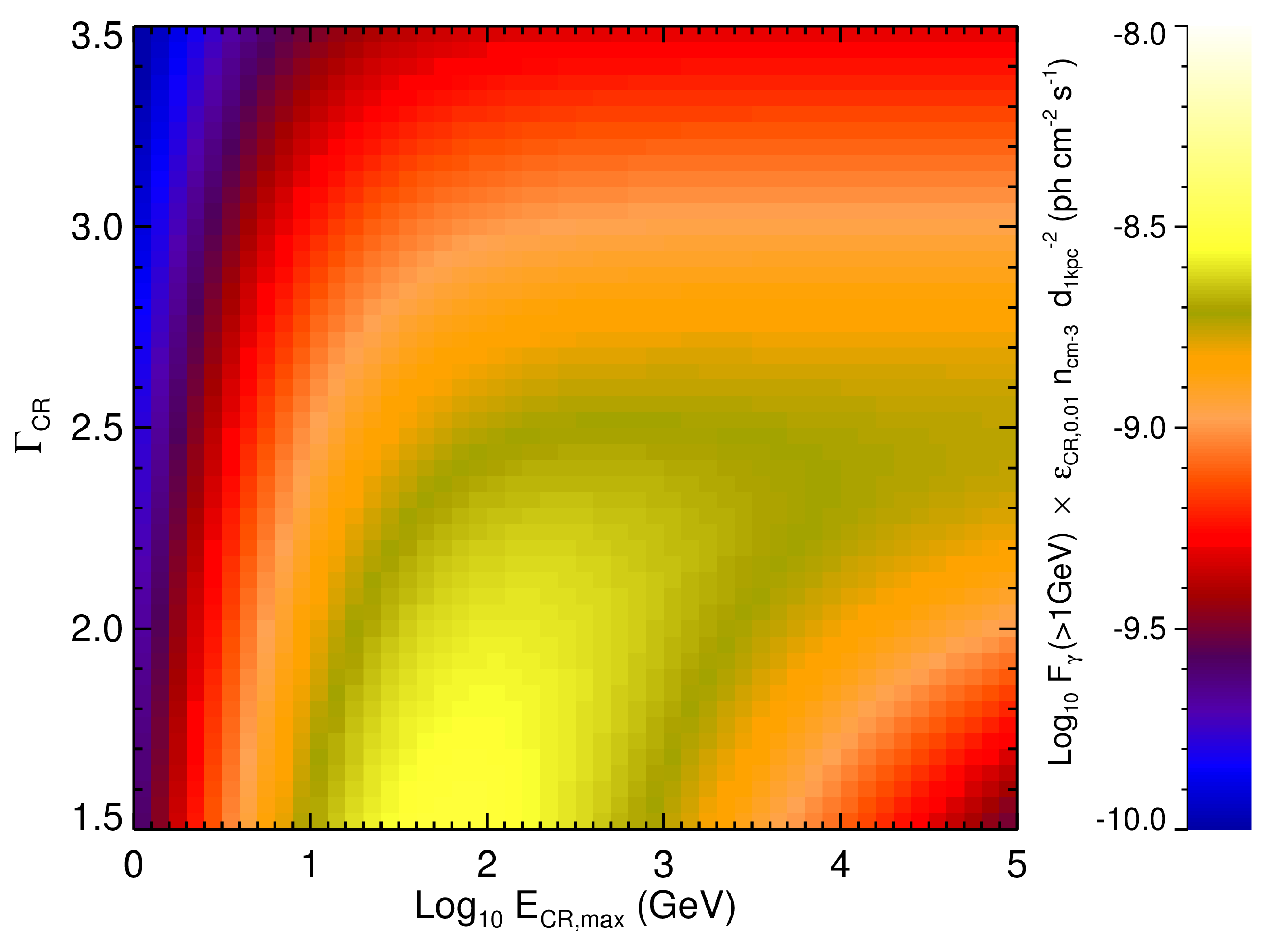}
  \caption{Under standard assumptions, a SNR's $\gamma$-ray flux above $1$\,GeV can be related to the accelerated CRs' maximal energy and index for a given acceleration efficiency ($0.01$), effective density ($1$\,cm$^{-3}$), and distance to the SNR ($1$\,kpc).}
  \label{fig:CR_EvsIndex}
 \end{figure}

Alternatively, we can allow the CR index and maximal energy to vary. Fixing the acceleration efficiency to a reasonable $\epsilon_{CR} = 1\%$, we can plot the $\gamma$-ray flux as a function of $\Gamma_{CR}$ and $E_{CR, max}$, as seen in Figure \ref{fig:CR_EvsIndex}. 
 
Figure \ref{fig:CR_EvsIndex} shows that if we know the CR index and maximal energy for a SNR with a given $\gamma$-ray flux, distance, and density, we can determine its CR acceleration efficiency and thereby, the amount of energy going into CRs. \cite{GRsFromSNRsCRs} compare theoretically calculated efficiencies, including time evolution of the SNR-interstellar material system, to measured GeV luminosities  for several detected SNRs.  With the $1^{st}$ {\em Fermi} SNR Catalog, we have measured fluxes or upper limits for all SNRs in the energy range $1-100$\,GeV \cite{JacksProceeding}. Moreover, we can fix the CR index to, for instance, the GeV $\gamma$-ray index, self-consistently measured with the flux. Finally we can, for instance, constrain $E_{CR, max}$ by relating it to either a SNR's measured break energy or to the maximum energy inferred from CRs interacting with the interstellar medium and creating the diffuse Galactic $\gamma$-ray background. \cite{CRsFromDiffuseFSympProc} and \cite{CRsFromDiffuseICRCProc} illustrate the extraction of CR parameters from the diffuse Galactic $\gamma$-ray background.
Combining the flux (upper limit), inferred index, and upper limit on CRs' maximum energy, we can place an upper limit on the energy transferred from a given supernova explosion to its CRs. Doing so for all known SNRs yields the total energy being transferred to CRs. If this is less than the observed total CR energy content, within the limits of our assumptions, another source must contribute to accelerating particles to CR energies. 

We can examine this explicitly for SNRs with GeV flux upper limits and distances. Assuming that they are merely faint rather than less energetic, we can use an index of $2.5$, about average for those observed so far. With a maximum CR energy of $E_{CR, max}\gtrsim 200$\,GeV, we find that the GeV flux is nearly independent of the energy (for indices $\Gamma \gtrsim 2.0$), and scales equation\,(\ref{eq:accelEfficiency}) by $1.5$. Solving for the efficiency,
\begin{equation}
\begin{split}\label{eq:accelEfficiencyULs}
 \frac{\epsilon_{CR}}{0.1}  & \times \frac{n}{1\,\textrm{cm$^{-3}$}}  \\ 
 & \approx \frac{F(1-100\,\textrm{GeV}) }{1.5 \times 10^{-8} \,\textrm{cm$^{2}$ s}} \times \left(\frac{d}{1\,\textrm{kpc}}\right)^{2}  %
 \end{split}
\end{equation}
for SNRs with a canonical energy of $10^{51}$\,ergs. We  note that if we use an observed flux, this efficiency is that for particles accelerated up to and including the moment of observation.

As it is also necessary to know the density of the medium with which the SNR interacts as well as the distance to the SNR, for our preliminary study, we turn to the $\sim 175$ SNRs detected in X-rays \cite{FerrandSafiHarb2012}, in search of those with thermal emission. The thermal X-ray emission from the shock-heated inter- and/or circumstellar mediums places reasonable constraints on the density. Likewise, most of these SNRs have a distance estimate. Densities may subsequently also be obtained from measurements such as IR emission from collisionally heated dust and hydrodynamics. 

We will explore methods and implications for constraining Galactic SNRs' contribution to the observed CRs by studying the efficiency, including using flux upper limits under the assumption of entirely hadronic processes. 

\section{Conclusions}
By examining correlations between SNRs' radio and GeV indexes, we observed that, while several known young SNRs tend to follow the expected radio-GeV index correlation for standard emission mechanisms, the majority do not. This challenges the previously sufficient models assumptions. In particular, we explored the hypothesis that the underlying particle population(s) may have a spectral break, correlating to a break in their emission spectrum. The GeV-TeV index correlation does in fact show that several SNRs have spectral breaks at or between GeV and TeV energies.  

The GeV index's evolution with SNR age suggests that the decreasing shock speed or decreasing maximum acceleration energy may cause the GeV index to tend to soften. Combining this with the observed luminosity differences also allows a scenario where the fainter, harder young SNRs are interacting with less dense material, while older, interacting SNRs have eg reached nearby local overdensities, such as molecular clouds. Multiwavelength information in the context of the $1^{st}$ {\em Fermi} SNR Catalog will help disentangle the effects of evolution and environment. 

We also explored a method for constraining SNR's contribution to the observed Galactic CR flux using the flux and index measured in the $1^{st}$ {\em Fermi} SNR Catalog, constraining the maximum CR energy using the diffuse Galactic $\gamma$-ray flux, for 
SNRs with known distances. Such a constraint may allow us to infer if another source population may be contributing to the measured Galactic CR flux or alternatively, if we need to refine the assumptions made.

\vspace*{0.5cm}
\footnotesize{{\bf Acknowledgment:}{ The \textit{Fermi} LAT Collaboration acknowledges generous ongoing support
from a number of agencies and institutes that have supported both the
development and the operation of the LAT as well as scientific data analysis.
These include the National Aeronautics and Space Administration and the
Department of Energy in the United States, the Commissariat \`a l'Energie Atomique
and the Centre National de la Recherche Scientifique / Institut National de Physique
Nucl\'eaire et de Physique des Particules in France, the Agenzia Spaziale Italiana
and the Istituto Nazionale di Fisica Nucleare in Italy, the Ministry of Education,
Culture, Sports, Science and Technology (MEXT), High Energy Accelerator Research
Organization (KEK) and Japan Aerospace Exploration Agency (JAXA) in Japan, and
the K.~A.~Wallenberg Foundation, the Swedish Research Council and the
Swedish National Space Board in Sweden.

Additional support for science analysis during the operations phase is gratefully
acknowledged from the Istituto Nazionale di Astrofisica in Italy and the Centre National d'\'Etudes Spatiales in France.
}}

\end{document}